\documentclass{mem}
\usepackage{natbib}\usepackage{txfonts}\usepackage{balance}
\usepackage{graphicx}
\usepackage[a4paper]{hyperref}
\idline{75}{282}
\begin{document}
\def\teff{$T\rm_{eff }$}
\def\kms{$\mathrm {km s}^{-1}$}

\title{Accretion discs, coronae and jets in~black~hole~binaries: prospects for  Simbol-X}

   \subtitle{}

\author{
J. \,Malzac\inst{1} 
          }

  \offprints{J. Malzac }

\institute{
Centre d'Etude Spatiale des Rayonnements, OMP, UPS, CNRS;  9 Avenue du Colonel Roche, 31028 Toulouse, France
\email{malzac@cesr.fr}
}

\authorrunning{Malzac }

\titlerunning{The disc/jet connection in black hole binaries}

\abstract{
The phenomenology of accretion disc, coronnae and jets in X-ray binaries  is rather well established. 
However the structure of the accretion flow in the various spectral states is still debated and  the connection between the hot flow and compact jet is far from being understood. Simbol-X should address these two important questions in several ways.  First, it will provide us with the capability of producing high sensivity, broad band  spectra and therefore constrain simultaneously the shape and luminosity of all  spectral components (iron line, reflection bump, thermal disc and comptonised emission) which in turn provides information on the geometry of the accretion flow. It will also determine the exact contribution of jets to the X-ray band both in bright and quiescent states. Finally it will shed new  lights on the underlying mechanisms triggering spectral state transitions by allowing us to follow in exquisite details the rapid spectral evolution and its correlation with the radio jet emission during  those transitions. 
}
\maketitle{}

\section{Spectral states and the structure of the accretion flow}\label{sec:truncdisc}

\begin{figure*}[t]
\includegraphics[width=0.55\textwidth]{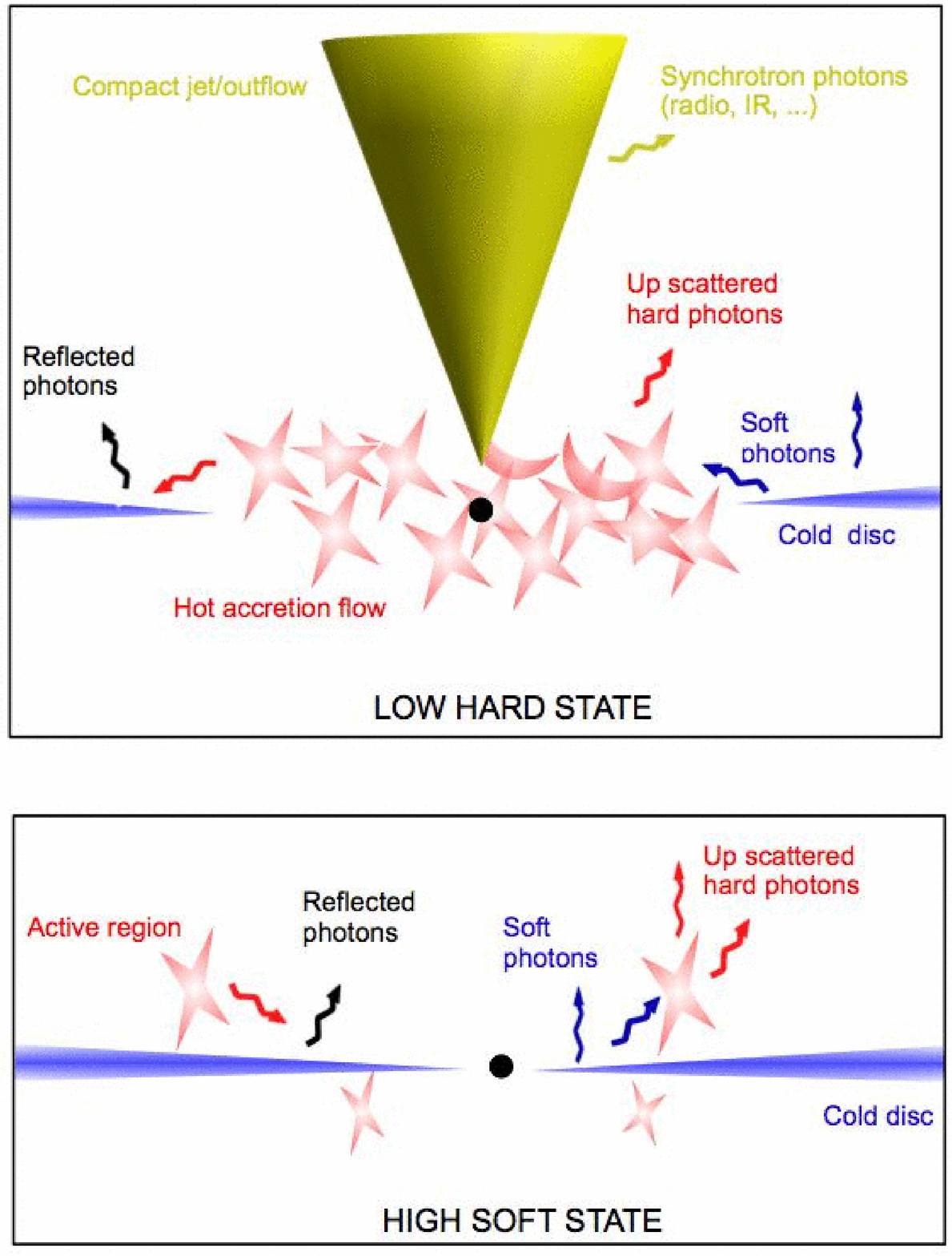}
\resizebox{0.5\textwidth}{!}{\parbox{9cm}{\vspace{-12cm}\hspace{+0cm}\includegraphics[width=8cm]{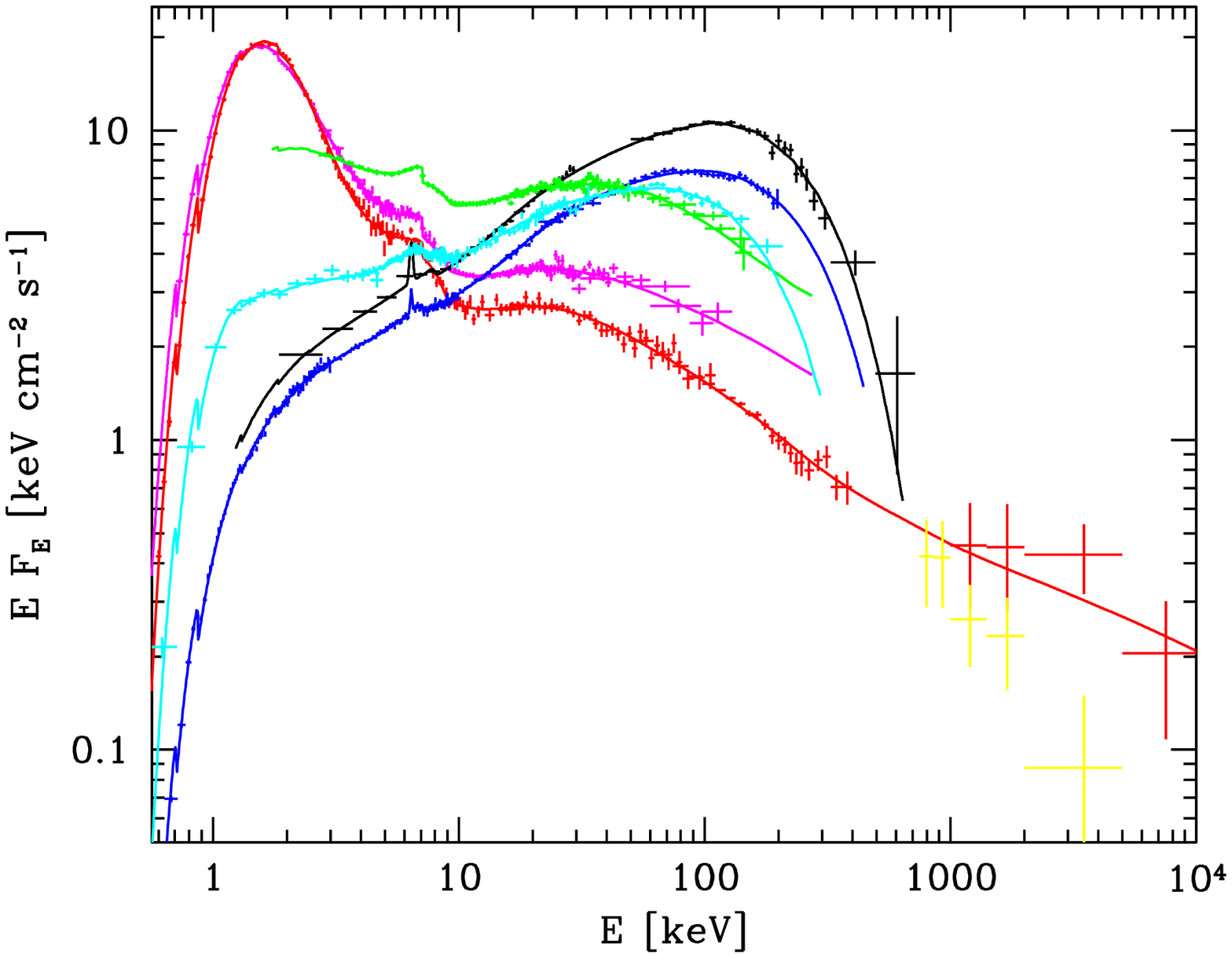}\\ \hspace*{0cm}\includegraphics[width=8cm]{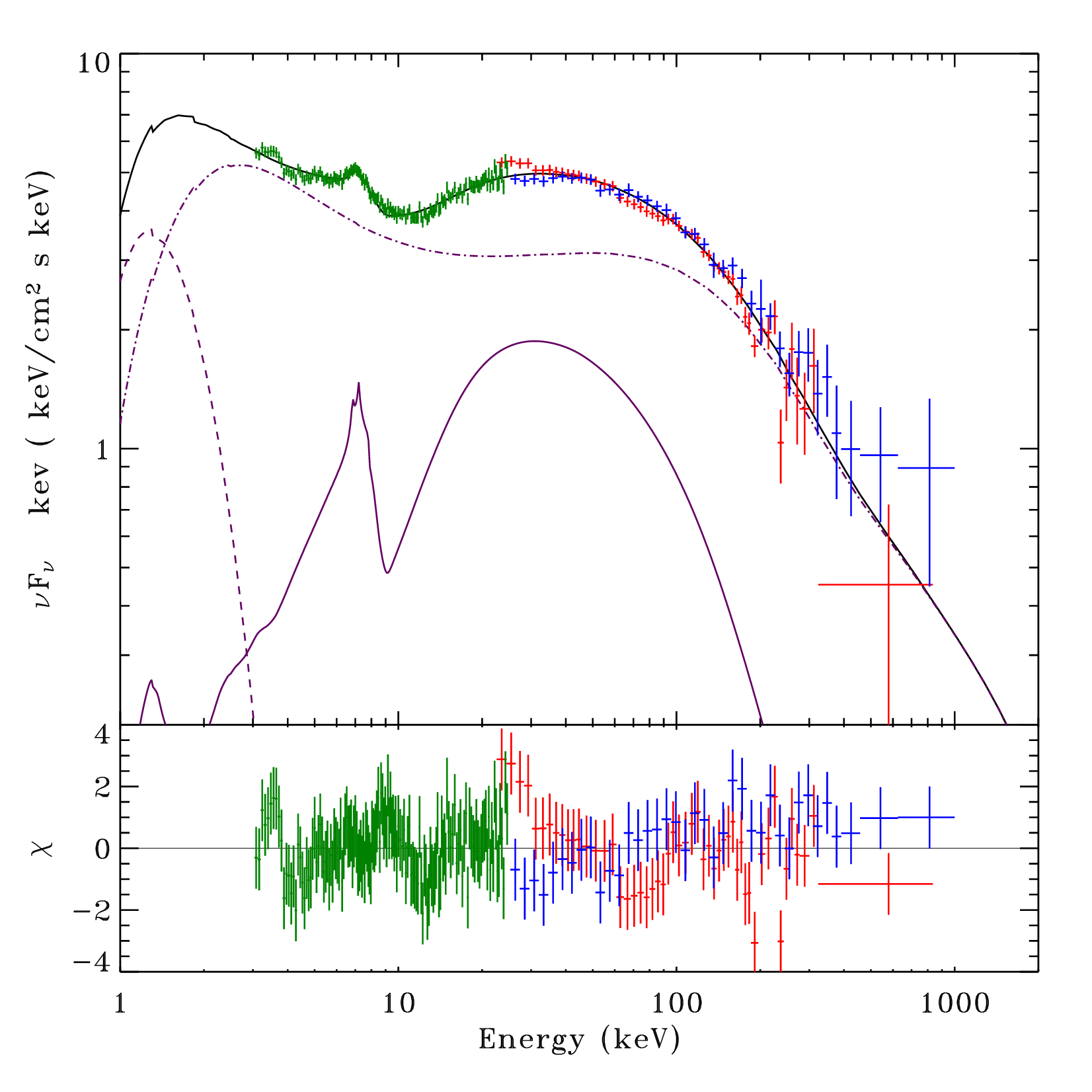}}}
\caption{Left hand side panels: Structure of the accretion flow during LHS (top) and HSS (bottom) according to the standard scenario. Right hand side panels: Observed spectra of Cygnus X-1 . Upper panel (from {Zdziarski et al 2002}):  the LHS (black and  blue), HSS (red, magenta), IMS (green, cyan). The solid curves give the best-fit Comptonization models (thermal in the hard state, and hybrid, thermal-nonthermal, in the other states). Lower panel:  Time averaged INTEGRAL spectrum of Cygnus X-1 during a mini-state transition (intermediate state).
 The data are fitted with the 
thermal/non-thermal hybrid Comptonisation model {\sc eqpair}
 with \emph{mono-energetic} injection of relativistic electrons. 
 The lighter curves show the reflection component (solid), 
 the disc thermal emission (dashed) and the Comptonised emission (dot-dashed).
 The green, red and blue crosses show the {\it JEM-X},
  {IBIS/ISGRI} and {SPI} data respectively. See {M06} for details.
\label{fig:cygx1spectra} }
\end{figure*}

Most of the luminosity of accreting black holes is emitted in the X-ray band. This X-ray emission is strongly variable.  A same source  can be  observed with very  different X-ray spectra (see e.g. 
 Done, Gierlinski \& Kubota 2007 for a recent review).  Fig.~\ref{fig:cygx1spectra} shows various spectra from the prototypical source  Cygnus~X-1 observed at different epochs. There are two main spectral states that are fairly steady and most frequently observed.
At high luminosities the accretion flow is in the High Soft State (HSS), caracterized by a strong thermal disc and reflection components (believed to be due to X-ray illumination of the accretion disc) and a weak and non-thermal component believed to be produced through  non-thermal  (or hybrid thermal/non-thermal) comptonisation in a hot corona. At luminosities lower than a few percent of Eddington ($L_{\rm Edd}$), the sources  are generally found in the Low Hard State (LHS) in which the disc blackbody and reflection features are much weaker, while  the corona has a thermal distribution of comptonising electrons and dominates the luminosity output of the system. Beside the LHS and HSS, there are several other spectral states that often appear, but not always, when the source is about to switch from one of the two main states to the other. Those states are more complex and difficult to define. We refer the reader to {McClintock and Remillard 2006} and {Belloni et al 2005} for two different  spectral classifications based on X-ray temporal as well as spectral criteria and radio emission.  In general, their spectral properties  are intermediate between those of the  LHS and HSS.

The different spectral states are usually understood in terms of changes in the geometry of the accretion flow. The standard picture is that in the HSS, there is a standard geometrically thin disc extending down to the last stable orbit and responsible for the dominant thermal emission. This disc is the source of soft seed photons for Comptonisation in small active coronal  regions located above an below the disc. 

  A population of  high energy electrons is formed through magnetic dissipation and  then cools down by up scattering the soft photons coming from the disc. This produces the high energy non-thermal emission which in turn illuminates the disc forming strong reflection features (see e.g. Zdziarski et al. 2002)

In the LHS, the standard geometrically thin disc does not extend to the last stable orbit, instead, the weakness of the thermal features suggest that it is truncated at distances ranging from a few hundreds to a few thousands gravitational radii from the black hole (typically  1000--10000 km). In these inner parts the accretion flow takes the form of a hot geometrically thick, optically thin accretion flow (Shapiro, Lightman and  Eardley 1976).  This hot flow is possibly  advection dominated (ADAF; see Ichimaru 1977; Narayan \& Yi 1994 )  and radiatively inefficient.
The electrons have a thermal distribution and cool down by Comptonisation of  the soft photons coming from the external geometrically thin disc, and IR-optical photons internally generated through self-absorbed synchrotron radiation.  

\begin{figure*}[t]
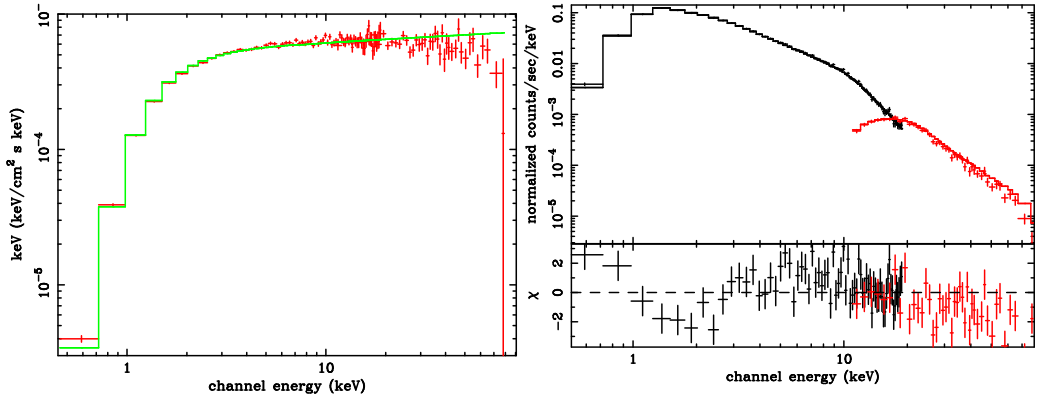

\includegraphics[angle=-90,width=0.5\textwidth]{malzac_f2a.ps}
\includegraphics[angle=-90,width=0.5\textwidth]{malzac_f2b.ps}
\caption{Simulation of a 400 ks observation of the quiecent source V404~Cyg  with the matrices provided for the conference. The shape of the observed Chandra spectrum (power-law with photon index $\Gamma=0.81$ absorbed by a column density N$_h$=0.7 10$^{22}$; Kong et al. 2002) was extrapolated into the hard X-rays domain assuming a high energy cut-off around 80 keV (cutoffpl model in XSPEC). The simulated data were then fitted with a simple absorbed powerlaw model.
The left hand side panel show the fit in  $\nu F_\nu$ representation while the right hand panel shows the count rates in the CZT and MPD detectors as well as the residuals.
The fit is poor (reduced $\chi^{2}$ of 1.94 for 136 degrees of freedom, null hypothesis probability 3 10$^{-10}$, F-test probability 2 10$^{-21}$). The slight curvature introduced by the high energy cut-of is therefore clearly detected. It can be seen in the residuals. }
\label{fig:v404cyg} 
\end{figure*}

However, the presence of the hot accretion disc is not the only possible model for the LHS. 
 It was suggested long ago that instead, a real accretion disc corona system (ADC) similar to that of the HSS may reproduce the spectra as well (Bisnovatyi-Kogan \& Blinikov 1976; Liang \& Price 1977) . This would imply a cold geometrically thin disc extending down very close to the black hole in the LHS. Unlike in the HSS, this disc does not produce a strong thermal component in the X-ray pectrum because it is too cold. It is cold because most of the accretion power is not dissipated in the disc, rather it is  transported away to power a strong corona and the compact jet. 
In this context the structure of the corona is tightly constrained by the observations (Haard \& Maraschi 1993). It must be patchy and consists of active regions of aspect ratios of order unity  outflowing with midldly relativistic velocities (Beloborodov 1999; Malzac, Beloborodov \& Poutanen 2001)

\section{The jet coronna connection}

Multi-wavelength observations
of accreting  black holes in the LHS have shown the presence of
an ubiquitous flat-spectrum radio emission which is absent in the HSS (see e.g Fender 2006 an references therein). 
The radio properties indicate it is produced
by synchrotron emission from relativistic electrons in compact,
self-absorbed jets (Hjellming \& Johnston 1988). 
Moreover, in the LHS sources a tight 
correlation has been found between the hard X-ray and radio luminosities, holding over more than three decades in luminosity 
(Gallo et al. 2003).

When the importance of the connection between radio and X-ray emission was realised,
it was proposed that the hard X-ray emission could be in fact synchrotron emission in the jet, rather  than  comptonisation in a hot accretion flow/corona (Markoff et al 2001). However, it seems that in most sources the synchrotron emission alone is not enough to reproduce the details of the X-ray spectra. In the most recent version of this model a thermal Comptonisation component was added which appears to provide a dominant contribution to the hard X-ray spectrum (Markoff et al 2005).  This component is supposedly located at the base of the jet which forms a hot plasma very similar to an ADC.  
There is however an observationnaly significant difference between the two models. Indeed, in the Markoff et al 2005 model, the jet synchrotron emission may provide a dominant contribution below $\sim$ 6 keV, while the Compton emission in the corona dominates at higher energies. As the synchrotron component is softer than the Comptonised one,  this feature fits quite well the spectral hardening  often observed around 5 - 10 keV.
In traditional Compton corona models, this hardening is ascribed to the effects of the reflection bump.  Its is also sometimes attributed to the effects of the soft thermal component from the accretion disc at low energy or the presence of a soft excess  (see e.g. Di Salvo et al. 2001). The possibility of performing broadband spectroscopy with Simbol-X (see Ferrando et al. in these proceedings), over an energy range (0.5-80 keV) covering both the thermal disc, the non-thermal corona (or jet) emission as well as the iron line and reflection bump, will certainly clarify the issue of the contribution of the jet to the hard X-ray emission of black hole binaries. 

In the context of ADC/hot disc models the correlation between X-ray and radio emissions, simply tells us that the corona and the compact jet of the LHS are intimately connected. 
A strong corona may be necessary to launch  a jet and/or could be  the physical location where the jet is accelerated or launched (Merloni and Fabian 2002). 
Jet spectral components have  recently been added to the ADAF model and make it possible to produce good fits of the whole spectral energy distribution from radio to hard X-rays (Yuan et al 2007 and references therein). In this model, the X-ray emission is almost always dominated by comptonised radiation from the ADAF.  However, an interesting prediction of this model is that  jet synchrotron emission could have a dominant contribution to the the X-ray emission of very faint sources (of luminosity lower than 10$^{-6}$ $L_{\rm Edd}$). This transition from an ADAF dominated to a jet dominated X-ray emission could be evidenced through  a change iin the  slope of the radio/X-ray  flux correlation at low luminosities (Yuan \& Cui 2005). This prediction could in principle be used to test this model. Gallo et al. (2006) have performed deep observations with the Very Large Array of A0620Ð00, leading to the first detection of radio emission from a black hole binary at X-ray luminosities as low as 10$^{-8.5}$ $L_{\rm Edd}$. Combined with simultaneous Chandra data, this observation  suggests that the non-linear correlation is maintained without change in slope down to quiecence. Actually, very  few sources with luminosities lower than 10$^{-6}$ are accessible to both radio and X-ray investigations.  An alternative approach could be to use X-ray spectroscopy alone. Indeed, at such low mass accretion rates an ADAF radiates essentially through Bremsstrahlung or inverse Compton in an optically (very) thin plasma. As a consequence,  the resulting X-ray spectrum should  be very bumpy.  On the other hand, jet synchrotron emission is expected to remain a power-law even at very low accretion rates. The spectroscopy of quiescent sources was performed with Chandra and XMM (Kong et al. 2002; Hameury et al. 2003) and usually,  in the limited bandwith of these instrument they show a powerlaw spectrum of photon index $\Gamma \sim 2$ (or harder) which is actually compatible with both a jet and ADAF interpretation. The high sensitivity of Simbol-X over a broader energy range, should be decisive in establishing the shape of the high energy spectrum of quiescent sources and  discriminating between those models. For example, the observation of a pure powerlaw emission extending up to 80 keV would certainly exclude an ADAF as the main source of hard X-rays in quiescent low mass X-ray binaries. A simulation of a Simbol-X  observation of the quiescent source V404 Cyg is shown in Fig.~\ref{fig:v404cyg}. This source is  a close (3.5 kpc) LMXB with quiescent  luminosities reaching 10$^{-6}$ $L_{\rm Edd}$. This simulation demonstrates that Simbol-X should be sensitive to even a small amount of curvature,  if any, in the quiecent spectrum of this source.

\begin{figure}
\includegraphics[ angle=-90,width=0.45\textwidth]{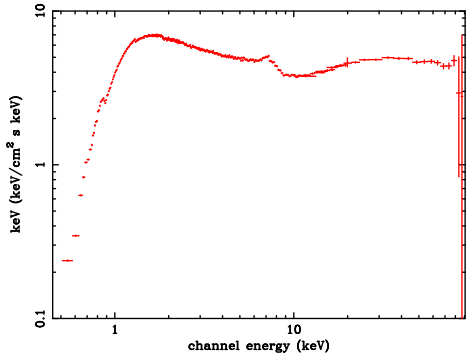}
\caption{ Simulated Simbol-X spectrum of a 1 ks observation of Cygnus X-1  with the matrices provided for the conference. These data were simulated according to the best fit EQPAIR model to the INTEGRAL spectrum  of the  2003 Intermediate state (Malzac et al. 2006) which is displayed on Fig.\ref{fig:cygx1spectra}.
\label{fig:cygx1simu} }
\end{figure}

Another interesting prospect for Simbol-X will be the possibility of performing time-resolved spectroscopy during states transitions. For instance Malzac et al. (2006; M06) report the results of an observation of Cygnus X-1 during 
a mini state transition. 
During this observation,  Cygnus X-1 was displaying spectral properties that were intermediate between the LHS and HSS and showing strong variability both in flux and spectral shape. 
The source was in an unstable transition state between LHS and HSS.
Broad band observations during those intermediate states are still rather rare and appear particularly interesting for the understanding of state transitions in BHBs.
 M06 produced light curves with a sampling time  of about 1 kilosecond in a dozen of relatively broad energy bands. They then applied a principal component analysis which demonstrated that most of the variability occurs through 2 independent modes. The first mode consists in changes in the overall luminosity on time-scales of hours with almost constant spectra (responsible for 68\% of the variance) that are strikingly uncorrelated with the variable radio flux. This variability mode was interpreted as variations of the dissipation rate in the corona, possibly associated with magnetic flares. The second variability mode consists in a pivoting of the spectrum around ~10 keV (27\% of the variance). It acts on a longer time-scale: initially soft, the spectrum hardens in the first part of the observation and then softens again. This pivoting pattern is strongly correlated with the radio (15 GHz) emission: radio fluxes are stronger when the INTEGRAL spectrum is harder. M06 proposed that the pivoting mode represents a ``mini" state transition from a nearly HSS to a nearly LHS, and back. The data are consistent with the  mini-transition being caused by changes in the soft cooling photons flux in the hot Comptonising plasma associated with an increase of the temperature of the accretion disc. The jet power then appears to be anti-correlated with the disc luminosity and unrelated to the coronal power. The reason for the anti-correlation between jet and X-ray luminosity is most probably  that state transition are associated with a redistribution
of the available accretion power between the
compact jet and the cold accretion disc. In the standard hot disc scenario described in Sec.~\ref{sec:truncdisc}  this redistribution of
accretion power could occurs because the jet shrinks as
the inner radius of the outer disc moves closer to the
black hole.
 From this interpretation we also infer that the bolometric luminosity jumps by a factor of about 2 during the transition hard to soft, suggesting a radiatively inefficient accretion flow in the Low Hard State. 

However,  this interpretation relies on the  assumption that the temperature, (or at least the  luminosity) of the Shakura-Sunyaev accretion disc  changes during the transition. Unfortunately,  this cannot be observed directly  with INTEGRAL since the thermal disc emission peaks around 1 keV and the INTEGRAL/JEM-X lower energy threshold is only at 3 KeV (see model spectrum and data on Fig.~\ref{fig:cygx1spectra}).  Other contemporary instruments, such as XMM would have been able to detect changes in the cold disc luminosity, but their lack of sensitivity above 10 keV would have prevented the identification the pivoting pattern of the Comptonised component. On the other hand, Simbol-X with its high sensitivity over a broad energy range  will certainly provide a critical test for  state transition models. The rapid evolution of the thermal disc emission, hard comptonised component and reflection features (both line and reflection component) will be followed simultaneously  and accurately. Fig.~\ref{fig:cygx1simu} shows that 1ks exposure will be enough to have  all the features of the the high energy spectrum of Cygnus X-1  revealed in exquisite details. Such broad band, time resolved spectroscopy studies of bright sources will certainly provide extremely valuable information on the evolution of the geometry and energetics of the accretion flow during spectral state transitions (see also Rodriguez et al. in these proceedings)

\bibliographystyle{aa}

\end{document}